\documentstyle[epsf]{mn}

\title[Turbulent MHD model for molecular clouds]
{A turbulent MHD model for molecular clouds and a new method of accretion
on to star-forming cores}

\author[D. Balsara, D. Ward-Thompson, R. M. Crutcher]
       {D. Balsara$^1$, D. Ward-Thompson$^2$, R. M. Crutcher$^{1,3}$ \\
        $^1$NCSA, University of Ilinois, Champaign-Urbana, Illinois, USA \\
        $^2$Dept of Physics and Astronomy, Cardiff University, 
            PO Box 913, Cardiff CF2 3YB \\
        $^3$Dept of Astronomy, University of Ilinois, Champaign-Urbana, 
            Illinois, USA}

\date{Accepted 2001 April 1; Received 2001 March 13; in original form 1999
December 1.}

\pagerange{\pageref{firstpage}--\pageref{lastpage}}
\pubyear{2001}

\begin{document}

\maketitle

\label{firstpage}

\begin{abstract}
We describe the results of a sequence of simulations of gravitational collapse
in a turbulent magnetized region. 
The parameters are chosen to be representative of molecular cloud material.
We find that several protostellar cores and
filamentary structures of higher than average density form. The filaments
inter-connect the high density cores. Furthermore, the magnetic field
strengths are found to correlate positively with the density, in agreement with
recent observations. We make synthetic channel maps
of the simulations and show that material accreting onto the
cores is channelled along the magnetized filamentary structures. This is 
compared with recent observations of S106, and shown to be consistent
with these data. We postulate that this mechanism of accretion along
filaments may provide a means for molecular cloud cores to grow to the
point where they become gravitationally unstable and collapse
to form stars.
\end{abstract}

\begin{keywords}
stars: formation
\end{keywords}

\section{Introduction}

Surveys of the densest cores of molecular clouds (eg: Benson \& Myers 1989;
Ward-Thompson et al. 1994) have shown them to be sites of active star 
formation, which can collapse under self-gravity to form objects known as 
protostars. A protostar is an object which will evolve into a star, but 
which is currently in the process of accreting the major part of its 
final main sequence mass (eg: Andr\'e, Ward-Thompson \& Barsony 1993 ;
Shu et al 1993).
The nature of this accretion process is still a 
matter of debate, although it is known that for low-mass stars 
(M $\leq$ 4--5 M$_\odot$) the protostar finally emerges from its 
enveloping cloud of material onto a well-defined pre-main sequence
track on the H-R diagram (Stahler, Shu \& Taam 1980), and hence to the
main sequence.

Much current debate centres around how dense cores in molecular 
clouds form initially, and how they subsequently evolve. One school of thought
suggests that cores form by Jeans-type gravitational instabilities (eg:
Blitz \& Williams 1997), and subsequently evolve to higher densities by means 
of ambipolar diffusion (eg: Ciolek \& Mouschovias 1994). This is a process
whereby a large-scale magnetic field threading the dense core supports the
ionised component of the material against collapse while the neutral gas
diffuses under self-gravity towards the centre of mass. This causes the centre
of the core to increase in density until a critical mass-to-flux ratio is
reached, and runaway gravitational collapse sets in.

However, there is a growing body of observational evidence that these
quasi-static equilibrium processes are not the only important physical 
processes
in the evolution of the interstellar medium. Excess emission at higher
velocities than purely thermal emission (e.g. Falgarone \& 
Phillips 1990) have been explained in terms of intermittent velocity
behaviour, which is a characteristic of turbulence. It has also been shown 
that the linewidths of molecular gas emission can be decomposed into a thermal
and a non-thermal component (e.g. Casselli \& Myers 1995), where the 
non-thermal component is produced by turbulence. Likewise, it has been 
seen (Crutcher 1999) that there is a strong tendency for
molecular cloud material to have velocities that are supersonic
and mildly sub-Alfvenic. Therefore, any complete
understanding of the physics of the interstellar medium should include
consideration of a self-gravitating magnetohydrodynamic (MHD) fluid 
undergoing turbulence. 

\begin{figure}
\setlength{\unitlength}{1mm}
\begin{picture}(60,120)
\includegraphics{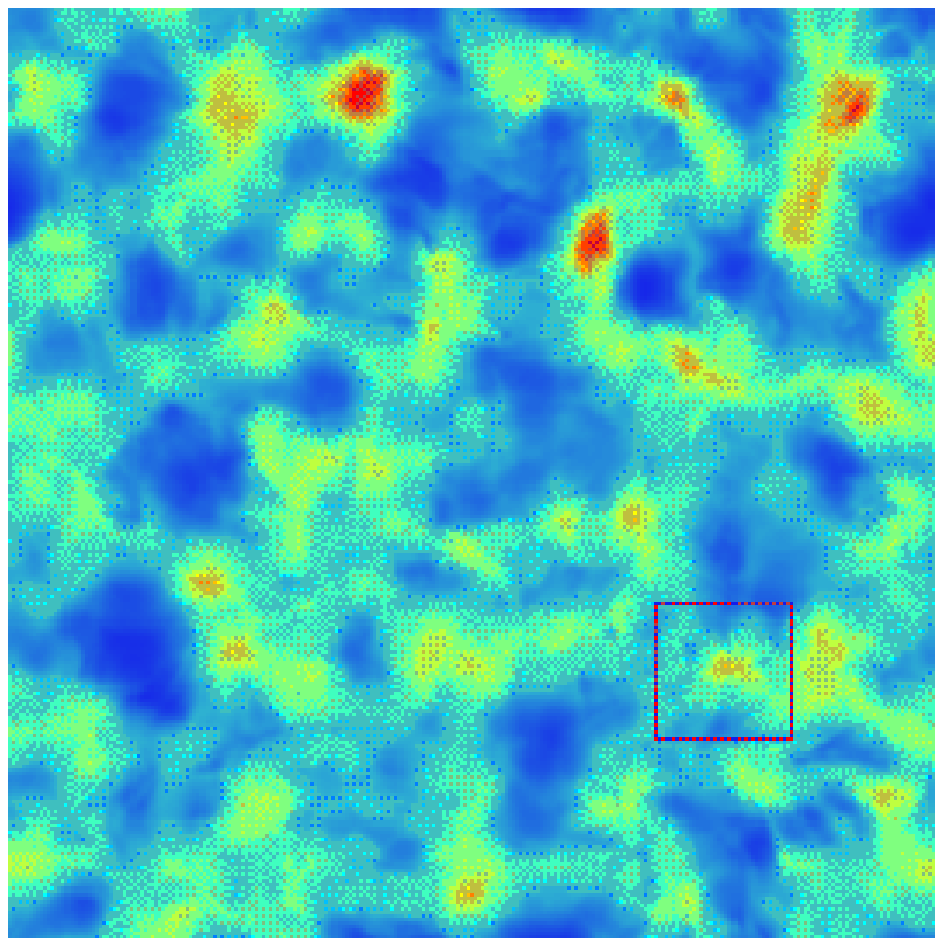}
\includegraphics{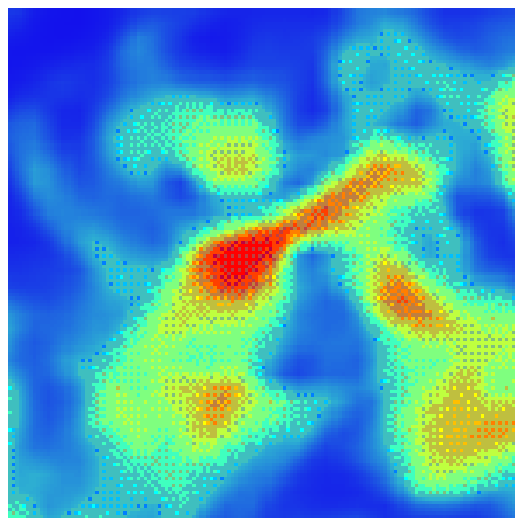}
\end{picture}
\caption{Panel (a) (upper) shows the simulated total intensity, 
assuming optically thin 
emission, from the model cube once pronounced cores have formed.
The red outline identifies the core magnified in Figure 1(b).
Panel (b) (lower) shows the integrated intensity from the core
delineated in Figure 1(a). Velocity channel maps for this core
are shown in Figure 3.}
\end{figure}

There have been a number of MHD models that have recently been published.
For example, Gammie \& Ostriker (1996) showed that MHD turbulence can
inhibit gravitational collapse. However, that model was limited to a slab
geometry for molecular clouds. Subsequently, Ostriker, Gammie \& Stone
(1999) presented the results of a 2.5-D MHD simulation, and showed that the
ratio of magnetic to thermal pressures in a molecular cloud provides an
indicator as to how the cloud will evolve to form stars. A number of
authors have shown that turbulence tends to decay on relatively short
time-scales compared to the life-times of molecular clouds (e.g. Stone
et al., 1998; MacLow 1999). However, there are many potential driving
mechanisms of turbulence that could counteract this decay, such as
stellar winds, jets and outflows, as well as large-scale motions such as 
shearing caused by Galactic differential rotation.
In this paper we present the results of fully three-dimensional simulations
of turbulent MHD flows in molecular clouds and compare them with observations.

\section{The Model}

\noindent
A three dimensional cubic computational domain 
with $256^3$ zones that is initially 2 pc across
was set up with a density of $10^{-20}$ g cm$^{-3}$. A tapered exponential
spectrum of velocity and magnetic fluctuations was initially used (see 
Balsara \& Pouquet 1999; Balsara, Crutcher \& Pouquet 1996; Balsara, 
Crutcher \& Pouquet 1999). The initial conditions consisted of flow
that had a turbulent root-mean square
(rms) Mach number of $1.5$ and an Alfven Mach
number of unity. These initial conditions are
consistent with the observations of several molecular 
clouds, particularly those `starless' clouds in which star formation
has not yet begun
(e.g. Crutcher 1999; Benson \& Myers 1989 and references therein). 
Periodic boundary 
conditions were utilized. The initial conditions are very similar
to the unforced models described in Balsara \& Pouquet (1999) with
the only difference that the simulation presented here is self-gravitating.

This simulation is one of several isothermal
and mildly adiabatic (adiabatic index of 1.2)
simulations that we have carried out over a range
of realistic turbulent rms Mach numbers and Alfven Mach numbers.
Truelove et al. (1997) found that isothermal collapse can form 
gravitationally condensed structures on all scales, which is
unphysical, but that an adiabatic equation of state can arrest collapse on
certain scales. A compromise to this issue was suggested by
Boss et al. (2000), who advocated using a barotropic equation of state.
We intend to explore this solution in subsequent papers, but we note that 
the compromise between adiabatic and isothermal equations of state
that we present herein appears to overcome the problems noted by
Truelove et al. (1997). 

The simulation was carried out on a fixed grid, and
it is well documented that the grid-based
criterion needs to be satisfied for the condensed objects (cores)
in order for their inner structure
to be properly analyzed (e.g. Truelove et al. 1997).
Satisfying the grid-based Jeans criterion is not always possible on
all scales. However, our focus in this
paper is not on the inner detailed structure of the cores, rather
it is on the accretion that takes place on to the cores.
For those length-scales our conclusions are unaffected
by the grid-based Jeans criterion.
We evolved the model using the RIEMANN code for
numerical MHD (Roe \& Balsara 1996; Balsara 1998a\&b;
Balsara \& Spicer 1999a\&b).

\begin{figure}
\setlength{\unitlength}{1mm}
\begin{picture}(60,50)
\includegraphics{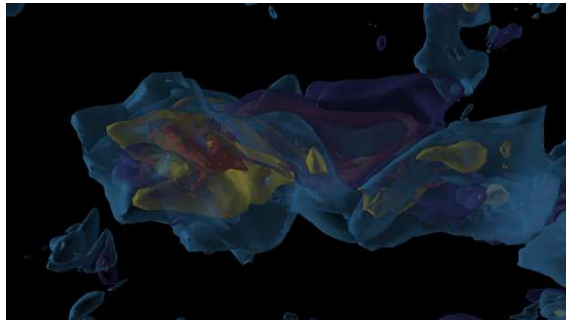}
\end{picture}
\caption{Isosurfaces of density and magnetic field magnitude superposed.
Cyan, yellow and red indicate isosurfaces of density that are four, six and 
eight times the mean density. Purple and magenta indicate isosurfaces of the
magnitude of the magnetic field that are five and seven times the mean.
Note the spatial coincidence of density and magnetic field strength.}
\end{figure}

\begin{figure*}
\setlength{\unitlength}{1mm}
\begin{picture}(230,230)
\includegraphics{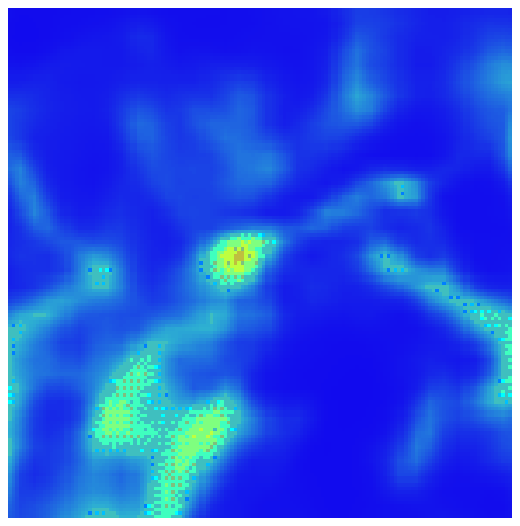}
\includegraphics{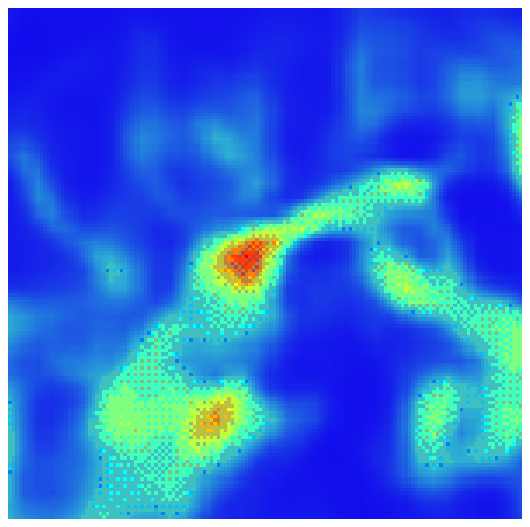}
\includegraphics{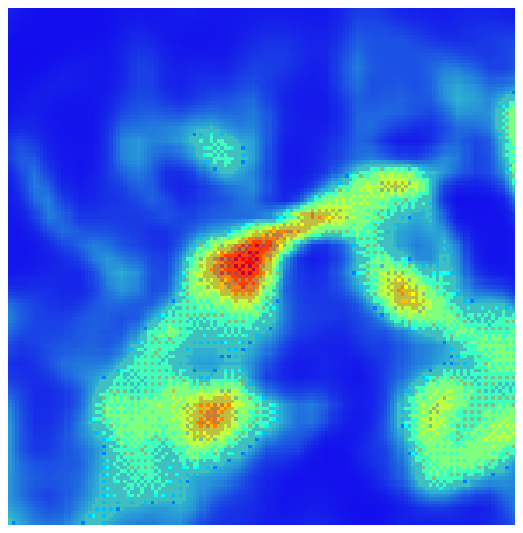}
\includegraphics{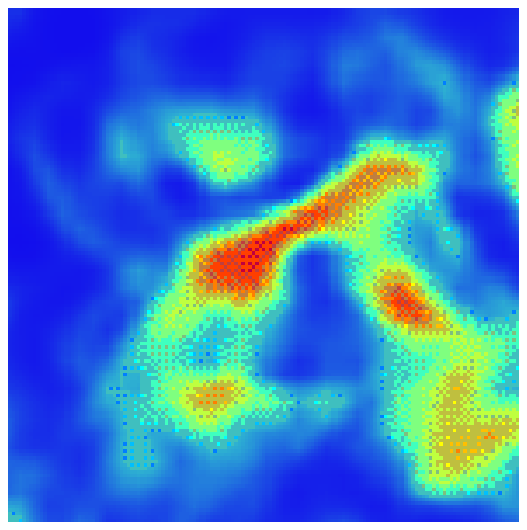}
\includegraphics{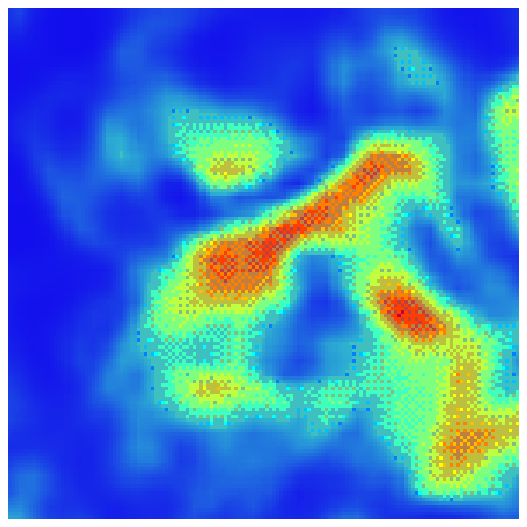}
\includegraphics{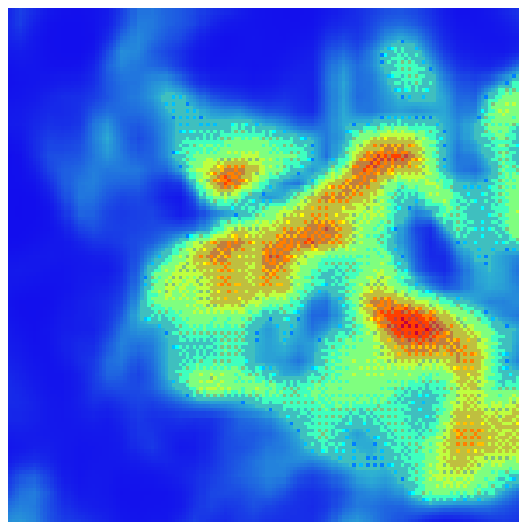}
\end{picture}
\caption{Panels (a)--(c) (left-hand column) and (d)--(f) (right-hand column)
show simulated channel maps of the same dense 
core shown in Figure 1(b) 
and its surrounding filamentary structure. Each panel 
represents a single velocity slice through the region. The 
ratios of the line of sight velocity to the thermal width
are $-$1.04, $-$0.80, $-$0.66, $-$0.41, $-$0.27 and $-$0.07 respectively.}
\end{figure*}

\subsection{Results of the simulations}

Figure 1(a) shows the simulated total intensity, 
assuming optically thin emission,
from the model cube after a few crossing times. 
It can be seen that the model molecular cloud has some regions of lower than
average density, as well as compact
regions of much higher density. Furthermore, the 
regions of highest density appear to be connected to one another by linear
features. Examination of the whole model cube reveals that these linear
features are essentially one-dimensional filaments, rather than 
two-dimensional sheets seen `edge-on'. All of the filaments terminate in at
least one high density region, while many of the filaments link two high
density regions. The filaments arise as a result of converging gas flows in
a turbulent medium.

We also see in our simulations that matter accretes on to
the cores along the filaments. The high density 
cores are all compact, and have
diameters ranging from a few hundredths of a parsec to little
over a tenth of a parsec. 
The larger cores are reasonably well-resolved,
occupying over twenty zones. However, all of the cores satisfy the
numerical resolution criterion (e.g. Truelove et al. 1997).
The densities
of the cores are high, having typical values of n(H$_2$)
$\geq$6$\times$10$^4$cm$^{-3}$. 
These sizes and densities are consistent 
with those observed in dense cores in actual molecular clouds (e.g. Benson \&
Myers 1989). 

We have analyzed the correlation between the magnetic 
field strength and the density in such simulations (Balsara et al. 1999)
where we showed that statistically there is a very strong positive correlation.
In Figure 2 we show the correlation visually using iso-surfaces of density
and magnetic field magnitude superposed on the same image of an isolated core. 
Thus cyan, yellow
and red indicate iso-surfaces of density that are four, six and eight 
times the mean density. Purple and magenta indicate iso-surfaces of the
magnitude of the magnetic 
field that are five and seven times the mean. We see that both the density and
magnetic field correlate positively and also that they are stretched out
in elongated filamentary structures. 
Furthermore, the magnetic field direction typically lies along the cores,
where it could serve to `funnel' gas onto cores (see below).

Quantitative analysis shows
that several of the cores are magnetically supercritical, in agreement with
the observational findings (Crutcher 1999). The model has, therefore, 
illustrated the topological structures of density and magnetic field
in turbulent magnetized molecular clouds and cores.

\subsection{Accretion onto cores}

Figure 3 shows simulated channel maps of one of the dense cores and its
surrounding filamentary structures. The channel maps were made assuming
optically thin emission.  The region of Figure 1(a) that we
have magnified in Figure 3 is shown by the red outline in Figure 1(a)
and is also shown in close-up in Figure 1(b).  
The channel maps were obtained by convolving the line of sight velocity
with the thermal broadening. Analysis of the channel maps shows that the
magnified region consists of two cores that are very well separated in
velocity space. We focus on one of them here. 

Figure 1(b) shows the integrated intensity of the selected core. 
Each of the panels of Figure 3 represents a 
single velocity slice through the region. The caption gives
the ratio of the line of sight velocity to the thermal width. The range of 
those numbers clearly shows that the motions are indeed supersonic. 
Figure 3(c) corresponds to the core's rest velocity. 
From Figure 1(b) we notice that there
is one filament heading in a north-easterly direction from the core.

In most situations it is reasonable to
expect that matter will accelerate
along one part of a filament and also decelerate along another part. 
Gravity should cause the mass to accelerate towards a core, while certain 
magnetic field topologies, as well as the presence of shocks or density 
gradients in the accreting material, would cause matter
to decelerate as it approaches a core. The
simulated channel maps allow us to dissect the acceleration or decelaration
of mass as it flows along the filaments. In general the filaments can be
curved, but a simple model that assumes linear filaments that make
contact with a core at one of their ends allows us to draw some simple
deductions on how to interpret channel maps. 

Figure 4(a) shows
a schematic diagram of a core that has two filaments with 
accreting matter that 
is decelerating towards the core.  A little reflection shows that,
if there is pure deceleration of accreting matter that is flowing in a linear 
filament towards the core,
the local maximum in the emission intensity should appear to
move away from the core as one steps away (in velocity space) from the 
core's rest velocity. This is schematically illustrated in 
Figure 4(b), 
where we show the motion of the intensity maxima in the channel maps.
Pure accceleration of matter in a linear filament should 
show the opposite trend in the channel maps. 
It should also be pointed out that constant velocity 
flow along a curved filament can also appear to be an acceleration or a 
deceleration depending on the filament's orientation relative to 
the observer. 

The schematic diagrams in Figure 4 allow one
to interpret the simulated channel maps in Figure 3.
Figure 3(a) shows only the core with a very faint suggestion of
filamentary structure. Figures 3(a)--(f), viewed in sequence, show 
the intensity in the filament become progressively stronger as the core 
itself becomes fainter. In Figure 3(e) the core is decidedly fainter than the
filament and in Figure 3(f) it is very much fainter than the filament.
By viewing Figures 3(a) through to 3(f) in sequence one also notices that
the intensity maximum in the filament moves away from the core in a
north-westerly direction traced out by the filament in the total intensity
map of Figure 1(b). In view of the discussion above, we interpret this as
deceleration of matter along the filament in the simulated channel maps
in Figure 3. 

From the range of velocities quoted in Figure 3 we
notice that the velocity ranges over an entire Mach number in this case.
We have carried out such imaging for numerous cores in the simulation
presented here as well as in numerous other simulations that we 
have carried out.
While a filamentary structure that is aligned along the line of sight will
not stand out prominently, we found that most cores have filamentary structures
around them. Larger cores show more filamentary structures, a fact that finds
a natural explanation in this scenario, as will be shown below. The velocities
that develop along these filaments correspond to a range of one or more 
Mach numbers.

\begin{figure}
\setlength{\unitlength}{1mm}
\begin{picture}(80,60)
\includegraphics{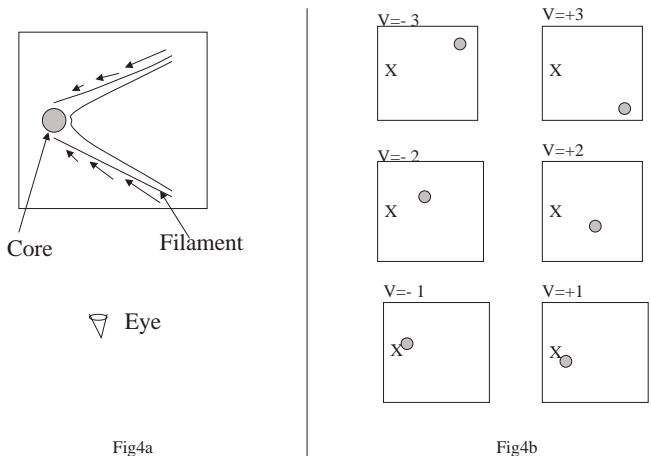}
\end{picture}
\caption{(a) Schematic diagram of a core undergoing accretion
along filaments. The shaded area denotes the core. The two linear
structures correspond to two filaments. The arrows along the filaments
denote velocity vectors. The rest velocity of the core is taken to be zero.
The image of the physical system is shown in plan view. To break the
degeneracy of having the two filaments appear in the same line we ask the
reader to tip the edge of the figure that is farther from the eye a little
above the plane of the paper. (b) Schematic diagram of the channel 
maps that would be produced by the situation in (a).
The `X' in the channel maps shows the core's location. The shaded blob
shows the location of the intensity maximum at that scaled velocity (which
is shown immediately above each channel map).}
\end{figure}

What we appear to be observing in Figure 3 is material decelerating
along a filament towards a dense core. Remember that the densest regions in 
the simulated molecular cloud are also the regions of highest magnetic flux
density. Thus the filaments, as well as being regions of higher density,
are also regions of higher magnetic flux density. Figure 2, as well as
our earlier results (see Balsara et al. 1998) have shown us that the 
magnetic field lines tend to lie along the filaments.

Hence these filaments are effectively magnetic flux tubes in the molecular
cloud, which are linked to the dense cloud cores.
So we interpret Figure 3 as illustrating how material can be channelled
by magnetic flux tubes and flow
onto the dense cores. Thus the model simulation has indicated one method in
which dense cores may evolve and grow in mass by accretion along magnetic
flux tubes. We also document that this process is seen to occur very often
for other cores that form in several of our simulations. 

\begin{figure*}
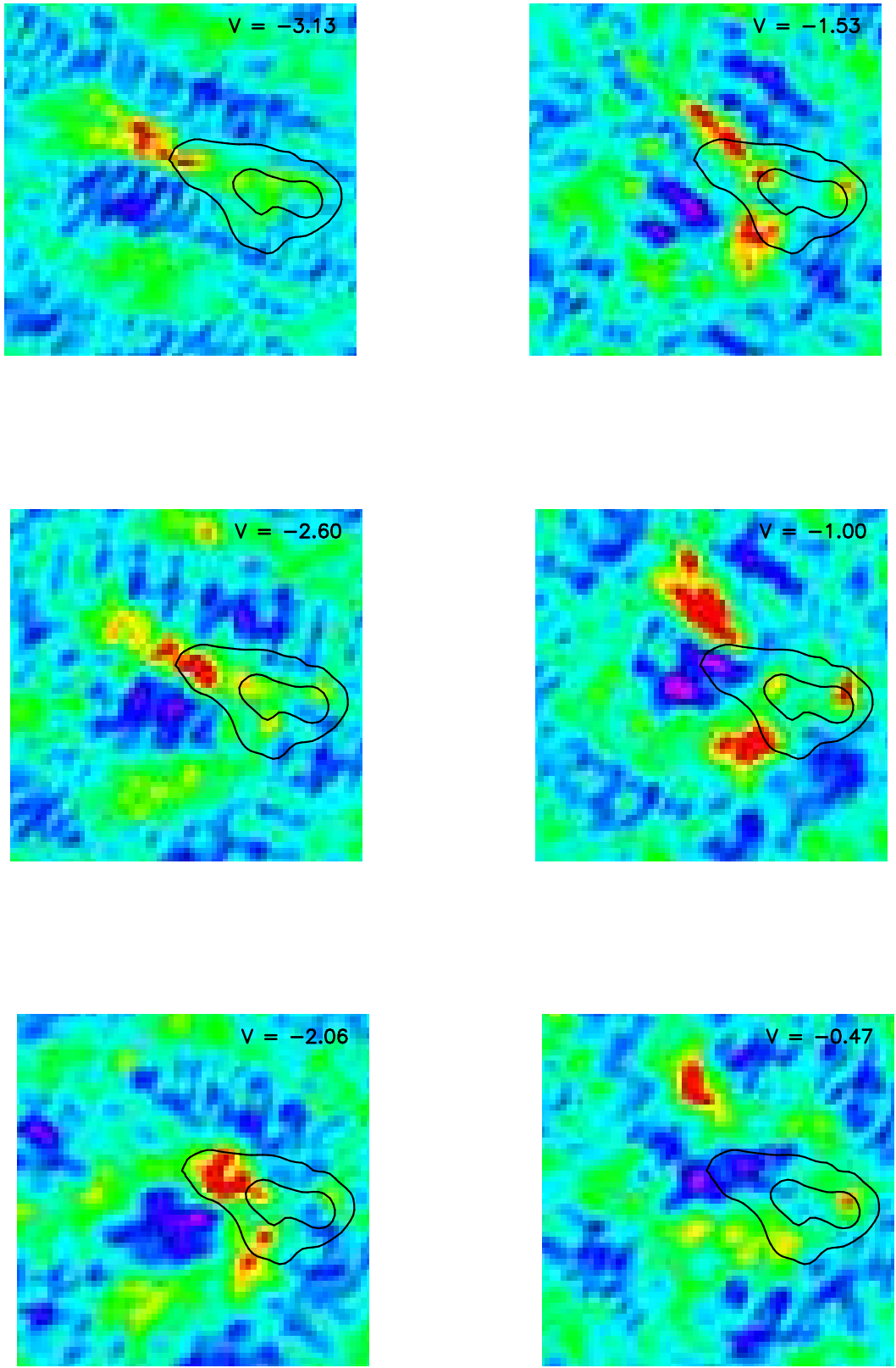

\setlength{\unitlength}{1mm}
\begin{picture}(230,230)
\includegraphics{fig5a.ps}
\includegraphics{fig5b.ps}
\includegraphics{fig5c.ps}
\includegraphics{fig5d.ps}
\includegraphics{fig5e.ps}
\includegraphics{fig5f.ps}
\end{picture}
\caption{$^{13}$CO channel maps of the S106 molecular cloud.
The black contours trace the 800-$\mu$m dust continuum emission and
therefore serve to identify the location of the protostellar core.
The numbers at the corners of the figures show the velocity in kms$^{-1}$.}
\end{figure*}

This suggests that the
amount of mass built up by a core over time will depend strongly on the 
topological structure of the magnetic fields that link to that core. The fact
that larger cores seem to have more filaments is now consistently explained
by the fact that larger cores are connected to more magnetic flux tubes which,
therefore, supply them with accreting matter at a faster rate. Our
new paradigm for core formation seems to form a substantial number of
magnetically supercritical cores. While consistent with recent observations
(e.g. Crutcher 1999) this fact is at variance with a previous paradigm for 
core formation (Mouschovias and Spitzer 1976) which suggests that cores that 
form are initially magnetically subcritical and only a portion of the
core becomes supercritical through the workings of ambipolar diffusion. 

The difference in the two paradigms stems from
the fact that matter is accreted directly along field lines in these
dynamical models while the previous models thought of core formation
as taking place through a sequence of quasi-steady states. In other 
work (Balsara,
Crutcher and Pouquet 1999) we showed that ambipolar diffusion does not
play a significant role in damping out the turbulence because the non-linear
eddy turnover times are shorter than the ambipolar diffusion time.
Hence the inclusion or exclusion of ambipolar diffusion does not strongly
affect these results.
In the next section we compare the model findings with one
particular molecular cloud region, to discover whether the predictions of the
model may actually occur in reality.

\section{Comparison with S106}

In Figure 5 we show $^{13}$CO channel maps of the S106 molecular cloud.
The black contours trace the 800-$\mu$m dust continuum emission and,
therefore, serve to identify the location of the protostellar core.
The numbers at the corner of the figures show the velocity in kms$^{-1}$
for each of the channel maps. Figure 5(d)
corresponds to the core's rest velocity. Based on temperature 
measurements of S106 the thermal sound speed can be estimated 
to be 0.5~kms$^{-1}$

S106 contains at its eastern edge S106-IR, and at its western edge the 
core which was identified from its sub-millimeter dust emission as S106-FIR 
(Richer et al. 1993). The latter source has no known outflow or associated 
protostar, and so is a candidate core for comparison with our model.
There are two filaments which join this core from 
the north-east that can be seen partly over-lapping in Figure 5. There is also
a much smaller filamentary structure that joins it from the south-east.
The two north-eastern filaments overlap each other to a great extent in the
total intensity maps but can be separated from each other in the channel maps.
The temperature in the filamentary gas has been measured and is too low
to permit one to interpret the filaments as outflowing gas.

The channel maps in Figure 5 of the velocity structure taken with the 
BIMA interferometer show evidence for velocity deceleration in the
gas that is accreting along the length of the filaments 
if one assumes that the filaments are not curved along the line
of sight (Roberts \& Crutcher 2001). However, one of the filaments 
in S106 appears to be curved in the plane of the sky, making it
difficult to deduce whether we are seeing acceleration or 
deceleration along it. In any case, the existence of filamentary 
structures and the motion of the intensity 
maxima along the filaments as one steps through the channel maps in Figure 5 
shows that these data are in close agreement with the 
model predictions shown in Figure 3. Thus we conclude that the accretion 
process in this core of S106 is taking place preferentially
along the filaments. 

Polarimetric observations (Hildebrand et al 1995; Holland et al 1996) suggest
that the magnetic field is also preferentially aligned along the long axis of
the filaments, a fact that is also consistent with the simulations. 
Thus the data for this cloud suggest a scenario where the magnetic 
fields serve to channel the accreting material along the filaments,
and the data are consistent with our hypothesised model scenario.
If this hypothesis is proved by further observations, then we have
shown how cores grow to become super-critical, and hence collapse
to form stars.

\section{Conclusions}

Using simulated channel maps from a numerical MHD simulation in conjunction
with observational data of S106, we have been able to
arrive at several insights into the process of accretion onto the molecular
cloud cores in which stars form:

(i) Turbulent MHD processes
cause the formation of high density cores;

(ii) The cores are linked by extended filamentary
structures;

(iii) Several of the simulated cores are found to be magnetically 
supercritical;

(iv) The higher density filaments are aligned with the
magnetic field structure;

(v) The accretion onto the
cores takes place along the filaments;

(vi) This suggests a
scenario where the material is
channelled by the magnetic field and flows under the gravitational influence
of the dense cores at the ends of the filaments, accreting onto
the cores themselves;

(vii) The amount of matter accreting on to a core thus
depends on the topological structure of the magnetic fields that link 
to that core;

(viii) Our observations of S106 appear to show this
process occurring in a real molecular cloud. 

If observations of other
regions show similar results, then this would show that we have distinguished
for the first time an important way in which star-forming cores accrete
matter to the point where they have gained sufficient mass to undergo 
dynamical collapse and form stars.

\section*{Acknowledgments}

DB wishes to thank A.Pouquet and C.McKee for useful discussions.
We also wish to thank S.Levy and R. Patterson for help with the images.
The use of SDSC supercomputer time is also gratefully acknowledged.

\label{lastpage}


\begin{thebibliography}{99}
\bibitem[]{}
Andr\'e P., Ward-Thompson D., Barsony M. 1993, ApJ 406, 122
\bibitem[]{}
Benson P. J., Myers P. C., 1989, ApJS, 71, 89 
\bibitem[]{}
Balsara D. S., Crutcher R. M., Pouquet A., 1996, in: Holt, S., Mundy L. G.,
eds., `Star Formation Near and Far', p. 89, AIP Press
\bibitem[]{}
Balsara D. S., Crutcher R. M., Pouquet A., 1999, ApJ, submitted
\bibitem[]{}
Balsara D. S., 1998a, ApJS, 116, 119
\bibitem[]{}
Balsara D. S., 1998b, ApJS, 116, 133
\bibitem[]{}
Balsara D. S., Pouquet A., 1999, Physics of Plasmas, 6, 89
\bibitem[]{}
Balsara D. S., Pouquet A., Ward-Thompson D., Crutcher R. M., 1999, in:
Franco J., Carraminana A., eds., `Interstellar Turbulence'
Cambridge Contemporary Astrophysics, p. 261
\bibitem[]{}
Balsara D. S., Spicer D. S., 1999a, J. Comp. Phys., 148, 133
\bibitem[]{}
Balsara D. S., Spicer D. S., 1999b, J. Comp. Phys., 149, 270
\bibitem[]{}
Benson P., Myers P. C., 1989, ApJS, 71, 89
\bibitem[]{}
Blitz L., Williams J. P., 1997, 488, L145
\bibitem[]{}
Boss A. P., Fisher R. T., Klein R. I., McKee C. F., 2000, ApJ, 528, 325
\bibitem[]{}
Caselli P., Myers P. C., 1995, ApJ, 446, 665
\bibitem[]{}
Ciolek G. E., Mouschovias T. Ch., 1994, ApJ, 425, 142
\bibitem[]{}
Crutcher, R. M., 1999, ApJ, 520, 706
\bibitem[]{}
Falgarone E., Phillips T. G., 1990, ApJ, 231, 438
\bibitem[]{}
Gammie C. F., Ostriker E. C., 1996, ApJ, 466, 814
\bibitem[]{}
Hildebrand R. H., Dotson J. L., Dowell C. D., Platt S. R., Schleuning D.,
Davidson J. A., Novak, G. 1995, in: Haas M. R., Davidson J. A., Erickson E. 
F., eds., `Airborne Astronomy Symposium on the Galactic Ecosystem'
ASP Conference Series, 73, 97
\bibitem[]{}
Holland W. S., Greaves J. S., Ward-Thompson D., Andr\'e P., 1996, 
A\&A, 309, 267
\bibitem[]{}
MacLow M., 1999, ApJ, 524, 169
\bibitem[]{}
Mouschovias T. Ch., Spitzer L. 1976, ApJ, 210, 326
\bibitem[]{}
Ostriker E. C., Gammie C. F., Stone J. M., 1999, ApJ, 513, 274
\bibitem[]{}
Richer J. S., Padman R., Ward-Thompson D., Hills R. E., Harris A. I., 
1993, MNRAS, 262, 839
\bibitem[]{}
Roberts, D., Crutcher, R. M. 2001, ApJ, submitted
\bibitem[]{}
Roe P. L., Balsara D. S., 1996, SIAM J. Appl. Math., 56, 57
\bibitem[]{}
Shu F. H., Najita J., Galli D., Ostriker E., Lizano S., 1993, 
in:  Levy E. H., Lunine J. I., eds., `Protostars and Planets III', 3, 
University of Arizona Press, Tucson
\bibitem[]{}
Stahler S., Shu F. H., Taam R. E. 1980, ApJ, 242, 226
\bibitem[]{}
Truelove J. K., Klein R. I., McKee C. F., Holliman J. H., Howell L. H.,
Greenough J. A., 1997, ApJ, 489, L179
\bibitem[]{}
Ward-Thompson D., Scott P., Hills R. E., Andr\'e P., 1994, MNRAS, 268, 276
\end{thebibliography}
\end{document}